\documentstyle[12pt]{article}
\makeatletter
\def\eqnarray{\stepcounter{equation}\let\@currentlabel=\theequation
\global\@eqnswtrue
\global\@eqcnt\z@\tabskip\@centering\let\\=\@eqncr
$$\halign to \displaywidth\bgroup\@eqnsel\hskip\@centering
  $\displaystyle\tabskip\z@{##}$&\global\@eqcnt\@ne
  \hfil$\displaystyle{\hbox{}##\hbox{}}$\hfil
  &\global\@eqcnt\tw@ $\displaystyle\tabskip\z@
  {##}$\hfil\tabskip\@centering&\llap{##}\tabskip\z@\cr}
\@addtoreset{equation}{section}
  \def\theequation{\thesection.\arabic{equation}}
\makeatother
\begin{document}
\date{\rightline{DTP/93/33}
\rightline{May, 1993}
}

\title{The Integrable Mapping as the Discrete Group of Inner Symmetry of
Integrable Systems}

\author{D.B. Fairlie\\ {\small \it University of Durham, South Road, Durham,
DH1 3LE, UK}\\
A. N. Leznov\thanks{On leave of absence from the Institute for
High Energy Physics, 142284 Protvino, Moscow Region, Russia} \\
{\small \it University of Durham, South Road, Durham, DH1 3LE, UK}}

\maketitle

\begin{abstract}
 It is shown that each integrable mapping is connected with a hierarchical
completely integrable sytem  of equations of
evolution type which are invariant with respect to the transformation
described by this mapping.
\end{abstract}

\section{Introduction}

Since the second half of the previous century it has been known that some
equations (or systems of equations) possess some nonlinear symmetries which
permit the generation of new solutions from known ones. These substitutions
are known as B\"acklund Transformations.The reader can find details of the
history of this subject in \cite{1}. Comparatively recently it has been
realised that such symmetries may be considered as a powerful method for the
construction of explicit solutions (including solitons, and doubly periodic
solutions) of all known completely integrable systems. The essence of this
construction consists in finding a suitable partial solution of the nonlinear
system under consideration and then in integration of its B\"acklund
transformation under the given boundary conditions. For the equations of
Self-Duality in four dimensions this approach was first employed in the work of
Fairlie, Corrigan, Goddard and Yates \cite{2}.For systems of more common
physical application, the explicit form of the B\"acklund transformation was
written
down and solved in \cite{3} The generalization to the case of equations in a
space of dimension 1+2 may be found in  \cite{4}.

This paper attempts to answer the the question `What conditions must a mapping
fulfil in order that it can be considered as the group of inner symmetries of
some integrable system of equations?' In other words we write down the system
of equations determining  the B\"acklund transformations of some integrable
system. We cannot as yet give the general solution - this procedure involves
functional equations, but in principle the general solution (or maybe
classification thereof) will provide the answer to the question of describing
all systems which possess a  B\"acklund transformation.

\section{ Condition of Integrability}

In what follows the term `integrable mapping' will play  the most important
role. The simplest way to understand this is from the point of view of integra
ble equations. Suppose we have an equation for a undetermined function $\theta$
in the form
\begin{equation}
F(\theta,\theta_i,\theta _{ij}, \theta_{ijk},\dots)=0
\label{1}
\end{equation}
where subscripts denote multiple differentiations with respect to the $N$
  independent variables $x_j,\ j=1,\dots, N$. Imagine that the solution to this
equation depends upon some parameter $\tau$ and differentiate the system
(\ref{1}) with respect to it., and call the result $\theta_\tau=\Theta$
\begin{equation}
{\partial F\over\partial\theta}\Theta+\sum{\partial
F\over\partial\theta_i}\Theta_i+\cdots =0.
\label{2}
\end{equation}
This is the symmetry equation for (\ref{1}). If some solution of the equation
of symmetry may be represented  in terms of $\theta$ and its derivatives up to
a fixed order, then equation (\ref{1}) is called integrable. In this case,
(\ref{1}) possess some form of exact solution. It may happen that (\ref{2}) may
admit a general solution. In this case (\ref{1}) is exactly integrable.
For example the universal equation proposed in \cite{4}   is of this kind. This
 equation in the simplest case in two dimensions is the so-called Bateman
equation
\begin{equation}
\theta_x^2\theta_{yy}-2\theta_x\theta_y\theta_{xy}+\theta_y^2\theta_{xx}=0.
\label{3}
\end{equation}

The symmetry equation for $\Theta$ is
\begin{equation}
2(\theta_x\theta_{yy}-\theta_y\theta_{xy})\Theta_x+2(\theta_y\theta_{xx}
-\theta_x\theta_{xy})\Theta_y+
\theta_x^2\Theta_{yy}-2\theta_x\theta_y\Theta_{xy}+\theta_y^2\Theta_{xx}=0.
\label{4}
\end{equation}
The general solution of this equation depending upon two arbitrary functions
may
be directly verified to take the form
\begin{equation}
\Theta= \theta_xF({\theta_y\over\theta_x})+G({\theta_y\over\theta_x})
\label{5}
\end{equation}
Thus the Bateman equation is exactly integrable, and its general solution
is well known to be given by the solution for $\theta$ of the implicit equation
\begin{equation}
 xf(\theta)+yg(\theta)=c
\label{6}
\end{equation}
where $f,g$ are arbitrary functions and $c$ is a constant, possibly zero.
The restriction to two dimensions is inessential.

\section{Discrete Integrable Substitution}

In the present section we shall consider local transformations acting on a
vector function of several independent arguments.  This means that this
substitution may be written down as
\begin{equation}
\tilde u=\varphi [u]\equiv \varphi (u,u_x,u_y,...u_{xx},u_{xy},u_{yy}...)
\label{6a}
\end{equation}
where the vector function $ \varphi $ depends on $ \vec u $ and its
derivatives up to order $ m $ taken at the same point. This restriction of
locality is not obligatory. The discrete symmetry transformation of the
equations of self duality in 4 dimensions is non-local, but all our
considerations also apply to this case.

The question is how to construct the system of equations which will be
invariant respect to the substitution (\ref{6}). In other words,
how to construct the dynamical system, the group of inner symmetry
of which corresponds to the substitution (\ref{6}); and to determine when
this is possible. We shall denote by  $ \vec \varphi_* $ the matrix
operator in the tangent space. Let us assume that the vector function $\vec u$
depends on some parameter $\tau$ and ask ourselves how the derivatives of
$\tilde u_\tau$ and $u_\tau$ are related to each other. The answer is
contained in the following definition of the operator  $ \vec \varphi_* $
\cite{ol} which acts as a matrix operator on the column vector $u_\tau$:
\begin{equation}
\tilde u_\tau =  \vec \varphi_* (u_\tau)
              = ( \vec \varphi_u + \vec \varphi_{u_i}D_i
             + \vec \varphi_{u_ij}D^2_{ij} +..... )(u_\tau)
\label{7}
\end{equation}
It is useful to rewrite (2) in a purely matrix form. Let us consider the
matrices $\varphi^{ij...k}_*$ with elements
$$
(\varphi_*)^\alpha\beta = \frac{\partial\varphi^\alpha}{\partial u_\beta}
  - \sum \partial^i \frac{\partial\varphi^\alpha}{\partial u^i_\beta}
  + \sum \partial^i \partial_j
            \frac{\partial\varphi^\alpha}{\partial u^{ij}_\beta}  - ....
$$
$$
(\varphi^i_*)^\alpha\beta = \frac{\partial\varphi^\alpha}{\partial u^i_\beta}
  - 2\sum \partial^j \frac{\partial\varphi^\alpha}{\partial u^{ij}_\beta}
  + 3\sum \partial^k \partial_j
            \frac{\partial\varphi^\alpha}{\partial u^{ijk}_\beta}  - ....
$$
$$
(\varphi^{ij}_*)^\alpha\beta = \frac{\partial\varphi^\alpha}{\partial
u^{ij}_\beta}
  - 6\sum \partial^k\frac{\partial\varphi^\alpha}{\partial u^{ijk}_\beta}
  + 10\sum \partial^k \partial_l
            \frac{\partial\varphi^\alpha}{\partial u^{ijkl}_\beta}  - .$$
and so on. In terms of this notation the relation (\ref{7}) may be written as:
\begin{equation}
\vec {\tilde F} (\varphi(u)) = \big( \varphi_* \vec F(u) \big)
    + \sum \partial_i (\big( \varphi^i_* \vec F(u) \big) )
    + \sum \partial_i\partial_j (\big( \varphi^{ij}_* \vec F(u) \big) )+...
,\label{8}
\end{equation}
where $\vec F(u) \equiv \vec u_\tau ,
\tilde {\vec F}(u) \equiv \tilde {\vec u}_\tau .$

{Definition}

The substitution (\ref{6}) will be called {\em integrable} if the vector
function
$\vec {\tilde F}$ in  (\ref{8}) as a function of its arguments coincides with
$\vec F$. If it is possible to find the general solution of (\ref{8}), then the
equation (\ref{6}) is exactly integrable.

If the substitution (\ref{6}) is {\em integrable}, then (\ref{8}) becomes the
equation  determining the vector function $\vec F$. From this construction
it is clear that each system of the form
\begin{equation}
\vec u_t = \vec F(u)
,\label{9} \end{equation}
where $\vec F(u)$ is any solution of (\ref{8}), is invariant under the discrete
substitution (\ref{6}).

The system (\ref{8}) is self consistent
 since for each substitution there exists
one obvious solution, viz., $\vec F(u) = D_i \vec u $, where $D_i$ denotes
differentiation with respect to the independent argument $x_i$. Now for
this solution, the system (\ref{9}) takes the simple form
$ u_t = u_{x_i}$, which is indeed manifestly invariant under every
substitution of the form (\ref{6}).

\section{Example of (1+2)d Davey Stewartson Equation}

We can take any of the known integrable systems, but here confine ourselves to
the case of the Davey Stewartson equation \cite{6} for the purposes of
illustration.
Consider the Toda lattice substitution in two
dimensional space which can be written in two equivalent forms:

$$
Sq \equiv \tilde q = {1\over r};
\quad Sr \equiv \tilde r = r(rq-(\ln r)_{xy})
$$
or
$$
 Su \equiv \tilde u = u - v_{xy} ;\quad  u=rq;\quad Sv \equiv \tilde v = v +
\ln \tilde u ;\quad v = \ln r$$
The general equation (3.3) in the variables $u,v$  takes the form:
$$
F^+_1 = F_1 - (F_2)_{xy} ,\quad
F_2 = F_2^- + {F_1 \over u} $$
where
$$
F^{\pm}_{\alpha} =  F_{\alpha}(S^{\pm1}u, S^{\pm1}v) $$
This system may be rewritten as a single equation for the  function
$F_1(u,v)=F$:
$$
F^+ - 2F + F^- = ({F \over u})_{xy} $$
or explicitly:
$$
F(u-v_{xy}, v+\ln(u-v_{xy})) - 2 F(uv) + F(u+(v-\ln u)_{xy}, v-\ln u)
= ({F(u,v)\over u})_{xy}
$$
We may directly check that apart from the obvious solutions $F = u_x , u_y$
the functional equation has the solution
$$
F= \alpha \partial_x (u_x - 2u v_x) +\beta \partial_y (u_y - 2u v_y)  $$
where $\alpha, \beta$ are arbitrary constants. For $F_2$ we immediately
obtain
$$
(F_2)_{xy}= \alpha [ - \partial_x \partial_y (v_{xx}+v_x^2) + 2u_{xx} ]
	     +\beta[ - \partial_x \partial_y (v_{yy}+v_y^2) + 2u_{yy} ] $$
which yields the system of two equations invariant with respect to
the above (Toda lattice) substitution:
$$
u_t =\alpha \partial_x (u_x - 2u v_x) +\beta \partial_y (u_y - 2u v_y)  $$
$$
 \partial_x \partial_y[v_t+ \alpha (v_{xx}+v_x^2) +
   \beta (v_{yy}+v_y^2) ]  = 2 (\alpha u_{xx}+ \beta u_{yy})  $$
This is the Davey-Stewartson equation. If we return to variables $r,q$, it
takes its original non-local form.

\section{Conclusion}

In the process of iteration of a discrete transformation of a given solution,
(say a mutltisoliton solution) to an integrable system both backwards and
forwards the general feature which arises is  that after a finite number of
steps the iteration becomes singular in either direction. An analogous
situation occurs in the construction of the general solution of
the $CP(N)$ models \cite{7} where the repeated action of the transformation
upon  an instanton solution produces the ani-instanton solution
after a finite number of steps. This situation is strongly reminiscent of what
happens in the representation theory of continuous groups; the finite
dimensional representations are obtained by the repeated action of ladder
operators on a highest weight state, and this action admits only a finite
number of repetitions before annihilation of the state vector. We are led to
conjecture that something analogous to a group action lies at the foundation of
all integrable systems, the properties of which we are only beginning to
discern.

\section{Acknowledgement}

A.N. Leznov wishes to thank the Royal Society for the award of a Kapitza
Fellowship which has enabled him to come to Durham and Thanks the Department of
Mathematical Sciences of the University of Durham for its hospitality.
\hfill\eject


\begin{thebibliography}{**}

\small

\bibitem{1}
Athorne C. and Nimmo J.J.C. {\it Inverse Problems} {\bf 7} 809  (1991).

\bibitem{2} Corrigan E.F., Fairlie D.B., Goddard P. and Yates R.G.  {\it
Communications in Mathematical Physics}~{\bf 58} 223-240  (1978) .

\bibitem{3} Leznov A. N., ``B\"acklund Transformation for Integrable Systems'',
IHEP
preprint 92--87 (Protvino, 1992).

 Leznov A. N., in {\it Proceedings of the First International A. D. Sakharov
Conference on Physics} (World Scientific, Singapore, 1991).

\bibitem{4} Leznov A.N., Shabat A.B. and  Yamilov R.I., {\it Physics Letters A}
(in Press)(1993).

\bibitem{5} Fairlie  D.B.,  Govaerts  J. and Morozov A, {\it Nuclear Physics B
373} 214-232  (1992).

 Fairlie  D.B.,Govaerts J. , Linearization of Universal Field Equations,
DTP-92/47to appear in {\it J. Phys A}  (1993) .

\bibitem{ol} Olver  P.J.{\it Applications of Lie Groups to Differential
Equations}
(Springer Verlag, 1986).

\bibitem{6} Davey A. and Stewartson K., {\it Proc. Roy. Soc. } {\bf A 338}
  101-110 (1974
).

\bibitem{7} Din  A.M. and  Zakrzewski W.J. {\it Nuclear Physics} {\bf B174}
W.J.397. (1980)

\end{thebibliography}
\end{document}